\documentclass{aastex62}

\newcommand{\Alfven}{Alfv\'{e}n }

\newcommand{\V}[1]{\mathbf{#1}}

\newcommand{\figref}[1]{Fig.~\ref{#1}}   
\newcommand{\secref}[1]{\S\ref{#1}}  
\newcommand{\blue}[1]{\textcolor{blue}{#1}}
  
\graphicspath{{./}{figures/}}

\usepackage[utf8]{inputenc}
\usepackage{amsmath}
\usepackage{soul}
\usepackage{graphicx}
\usepackage{amsmath}
\usepackage{amssymb}
\usepackage{verbatim}
\usepackage{xcolor}
\usepackage{url}
\usepackage{comment}
\usepackage{bm}

\received{}
\revised{}
\accepted{}
\submitjournal{ApJ}

\newcommand{\ssl}{\affiliation{Space Sciences Laboratory,University of California, Berkeley, CA 94720, USA}}
\newcommand{\pucb}{\affiliation{Physics Department,University of California, Berkeley, CA 94720, USA}}

\newcommand{\sao}{\affiliation{Smithsonian Astrophysical Observatory, Cambridge, MA 02138, USA}}
\newcommand{\umich}{\affiliation{Climate and Space Sciences and Engineering, University of Michigan, Ann Arbor, MI 48109, USA}}

\newcommand{\cnrs}{\affiliation{LPC2E, CNRS and University of Orléans, Orléans, France}}

\newcommand{\gsfchpl}{\affiliation{Code 672, NASA, Goddard Space Flight Center, Greenbelt, MD 20771, USA}}
\newcommand{\uofa}{\affiliation{Lunar and Planetary Laboratory, University of Arizona, Tucson, AZ 85721, USA}}

\newcommand{\swri}{\affiliation{Space Science and Engineering, Southwest Research Institute, San Antonio, TX 78238, USA}}
\newcommand{\unh}{\affiliation{Physics Department, University of New Hampshire, Durham NH 03824, USA}}
\newcommand{\mssl}{\affiliation{Mullard Space Science Laboratory, University College London, Dorking, RH5 6NT, UK}}
\newcommand{\unhssc}{\affiliation{Space Science Center, University of New Hampshire, Durham, NH 03824, USA}}

\shortauthors{Verniero et al.}

\begin{document}

\title{Strong perpendicular velocity-space in proton beams observed by Parker Solar Probe}

\correspondingauthor{Jaye Verniero}
\email{j.l.verniero@nasa.gov}

\author[0000-0003-1138-652X]{J. L. Verniero}\gsfchpl
\author[0000-0003-4177-3328]{B. D. G. Chandran} \unh
\author[0000-0001-5030-6030]{D. E. Larson}\ssl
\author[0000-0002-5699-090X]{K. Paulson}\sao
\author[0000-0001-6673-3432]{B. L. Alterman}\swri
\author{S. Badman}\ssl\pucb
\author[0000-0002-1989-3596]{S. D. Bale}\ssl\pucb
\author[0000-0002-0675-7907]{J. W. Bonnell}\ssl
\author[0000-0002-4625-3332]{T. A. Bowen}\ssl
\author[0000-0002-4401-0943]{T. Dudok de Wit}\cnrs
\author[0000-0002-7077-930X]{J. C. Kasper}\umich\sao
\author[0000-0001-6038-1923]{K. G. Klein}\uofa
\author[0000-0003-1945-8460]{E. Lichko}\uofa
\author[0000-0002-0396-0547]{R. Livi}\ssl
\author[0000-0001-6077-4145]{M. D. McManus}\ssl\pucb
\author[0000-0003-0519-6498]{A. Rahmati}\ssl
\author[0000-0002-0497-1096]{D. Verscharen}\mssl\unhssc
\author{J. Walters}\uofa
\author[0000-0002-7287-5098]{P. L. Whittlesey}\ssl
%
%
%
%




\begin{abstract}

\indent The SWEAP instrument suite on Parker Solar Probe (PSP) has detected numerous proton beams associated with coherent, circularly polarized, ion-scale waves observed by PSP's FIELDS instrument suite. Measurements during PSP Encounters 4-8 revealed pronounced complex shapes in the proton velocity distribution functions (VDFs), in which the tip of the beam undergoes strong perpendicular diffusion, resulting in VDF level contours that resemble a `hammerhead.' We refer to these proton beams, with their attendant `hammerhead' features, as the ion strahl. We present an example of these observations occurring simultaneously with a 7-hour ion-scale wave storm and show results from a preliminary attempt at quantifying the occurrence of ion-strahl broadening through 3-component ion-VDF fitting. We also provide a possible explanation of the ion perpendicular scattering based on quasilinear theory and the resonant scattering of beam ions by parallel-propagating, right circularly polarized, fast-magnetosonic/whistler waves.

\end{abstract}



\section{Introduction}\label{sec:int}

\indent Parker Solar Probe (PSP) \citep{Fox:2016} is the nearest human-made object to the Sun, making unprecedented observations of solar wind plasma and electromagnetic fields with every orbit. Prior to PSP, {\em Helios} was the spacecraft that had approached closest to the Sun, reaching a heliocentric distance $r$ of $\simeq 60 R_{\odot}$ or $\simeq 0.3 \mbox{ au}$. 
A series of Venus flybys is repeatedly modifying PSP's orbit, reducing the heliocentric distance of PSP's perihelion. PSP's most recent 9th perihelion on 2021-08-09 at $r\approx16 R_{\odot}$. Within a few years, PSP  will reach heliocentric distances just under $10 R_{\odot}$.
In this paper, we present new PSP observations of proton beams that undergo strong velocity-space diffusion perpendicular to the magnetic field, as well as a theoretical model that explains the shapes of the observed
velocity distribution functions (VDFs). 

Ion beams in the solar wind at $r> 0.3 \mbox{ au}$ have been studied extensively over the past five decades \citep{Feldman:1973,Feldman:1974,Marsch:1982,Marsch:1987,Neugebauer:1996,Steinberg:1996,Kasper:2006,Podesta:2011a,Klein:2018,Alterman:2018,Durovcova:2019}. These non-thermal features in the ion VDFs arise in part because of  the nearly collisionless nature of the solar wind. The free energy carried by the beams can drive plasma instabilities and provide an important source of plasma heating in the solar wind 
\citep{Daughton:1998,Hu:1999,Gary:2000,Marsch:2006,Maneva:2013,Verscharen:2013a,Verscharen:2013b,Verscharen:2015}. PSP's near-Sun observations provide unique insight at distances $r < 0.3$ au, where beams may be generated.

The secondary proton beam population in VDFs has now been recognized to play as significant role in energy transfer in the inner heliosphere as the core proton population. Joint observations by PSP's Solar Wind Electrons Alphas and Protons (SWEAP) \citep{Kasper:2016} and FIELDS \citep{Bale:2016} instrument suites show the presence of ubiquitous proton beams concurrent with ion-scale waves \citep{Bale:2019,Bowen:2020a,Verniero:2020,Klein:2021}.
PSP's first perihelion encounter with the Sun revealed a diversity of stream characteristics in the near-Sun solar wind, including large-amplitude Alfv\'enic ``switchbacks" and intervening ``quiescent" periods of near-radial magnetic field \citep{Bale:2019}. The quiescent solar wind in Encounters 4-5 illuminated striking patterns between features in the ion VDFs and wave activity. In particular, the proton beam tended to broaden (perpendicular to the magnetic field) over a relatively narrow energy range during periods of coherent ion-scale waves that are right circularly polarized in the spacecraft frame, producing VDFs with level contours with a `hammerhead' shape. The proton VDF populations that lie outside the core population contribute to the main asymmetries leading to energy transfer. Since these particles carry information regarding ion heat flux, we therefore define the \textit{ion-strahl} as the superposition of the narrow field-aligned beam population and the perpendicularly broadened hammerhead population.

In Section~\ref{sec:obs}, we describe a procedure for modeling the ion VDFs as a superposition of three particle populations: a core, a beam of nearly field-aligned particles, and a broadened `hammerhead' population. Using this procedure, we explore correlations between different properties of the hammerhead particles, as well as the association between the prevalence of hammerhead particles and the energy density of right-circularly polarized waves at the ion scale. In Section~\ref{sec:theory}, we develop a theoretical model for explaining the shapes of the hammerhead VDFs observed by PSP, and we test this model by comparing its predictions with PSP observations. We present our conclusions in Section~\ref{sec:conc}.

\section{Observations: The Hammerhead Feature as a new asymmetry in Proton VDFs}
\label{sec:obs}
 \begin{figure}[b]
\hspace{.45in} (a) Proton VDF \hspace{1.5in} 
\vfill
{\includegraphics[trim = 0mm 0mm 0mm 0mm, clip, width=0.35\textwidth]{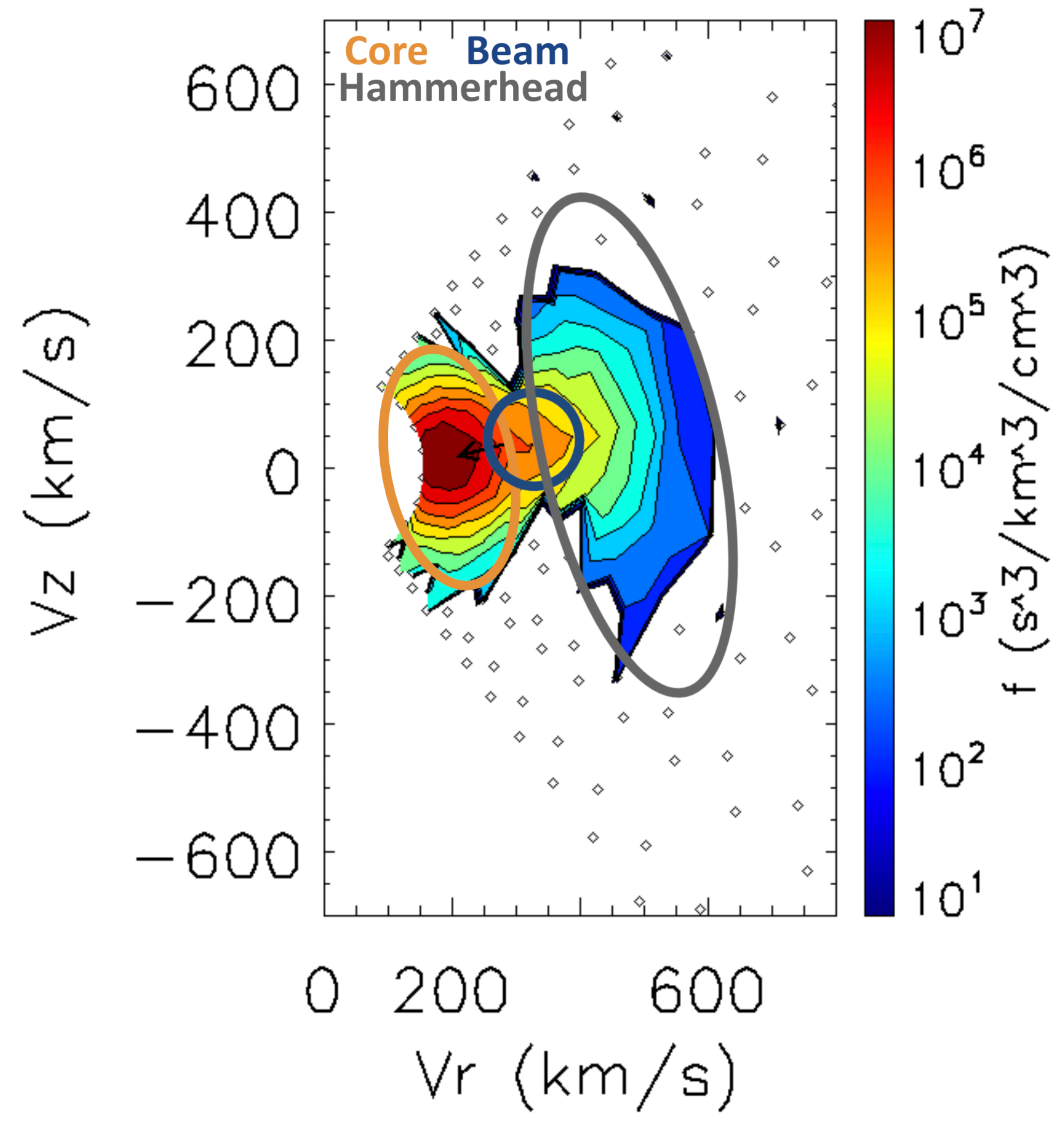}}
\hspace{.05in}
{\includegraphics[trim = 0mm 0mm 0mm 0mm, clip, width=0.65\textwidth]{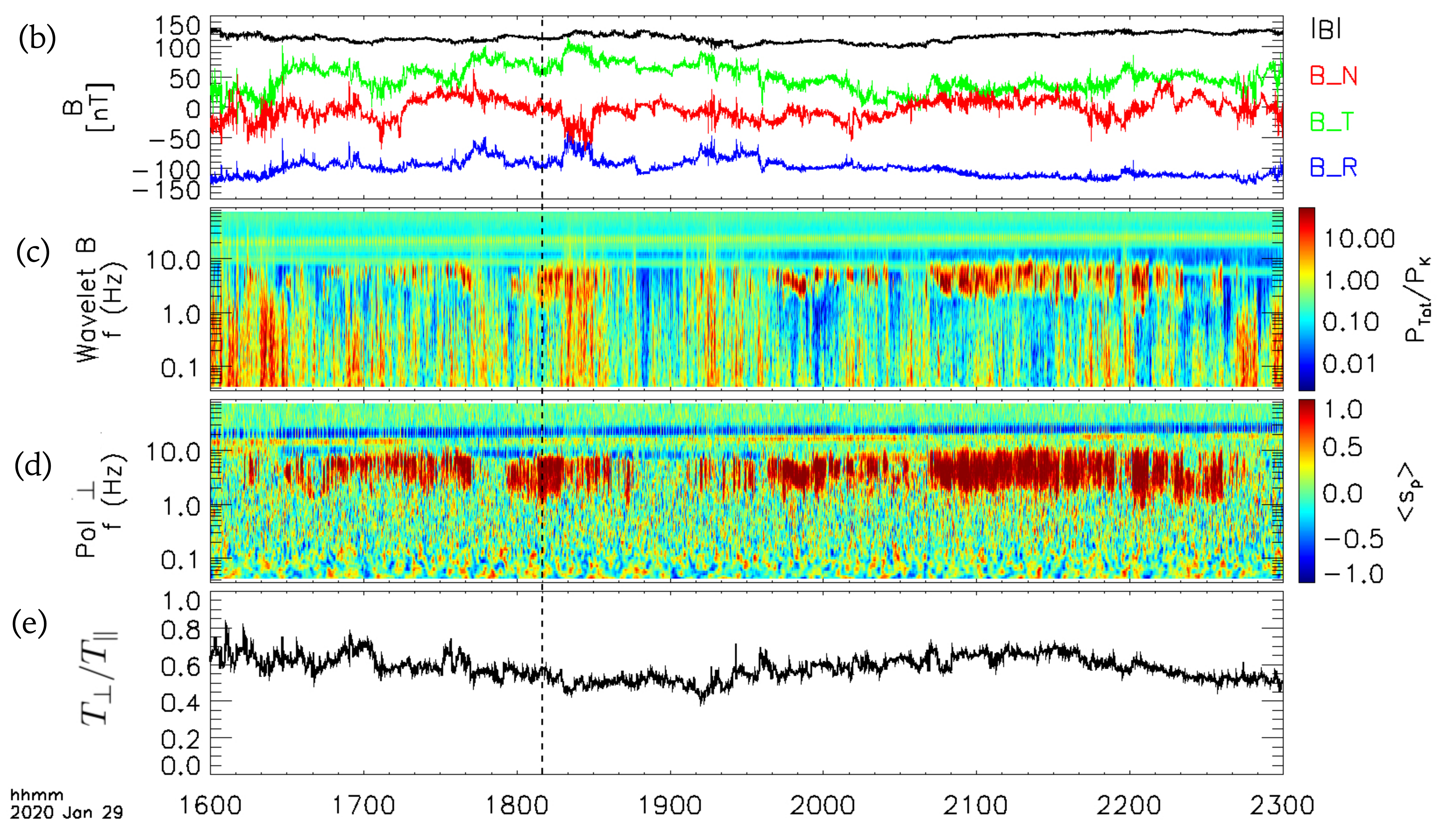}}
\caption{Left: (a) Example proton VDF in SPAN-I instrument coordinates during E4 quiescent solar wind period at vertical dashed black line shown on the right. The core, beam, and hammerhead populations are circled in orange, blue, and grey, respectively. The VDF shown was collapsed onto the $\theta$-plane in $(v_r,v_z)$ SPAN-I instrument coordinates, where $v_r = \sqrt{v_x^2+v_y^2}$. The black arrow indicates the direction of the magnetic field, rotated to SPAN-I instrument coordinates; the head of the arrow is placed at the solar wind velocity, with length representing the local \Alfven speed. Right: The FIELDS magnetometer observations show (b) $\V{B}$ in RTN coordinates, (c) Wavelet spectrogram of $\V{B}$ (with color scale as total power $P_{tot}$ normalized by background $k^{(-5/3)}$ $P_k$ power), (d) Perpendicular polarization of $\V{B}$ (red = RH, blue = LH in s/c frame). Panel (e) shows the SPAN-I measured temperature anisotropy, $T_\perp/T_\parallel$.
\label{fig:cbs}}
\end{figure}

Here we describe the PSP observations and methods used for data analysis regarding the prevalence of the ion-strahl broadening simultaneous with ion-scale waves. We combine measurements from both the Solar Wind Electrons Alphas and Protons (SWEAP) \citep{Kasper:2016} and FIELDS \citep{Bale:2016} instrument suite. 
SWEAP combines measurements from a sun-pointed Farday cup (FC) called the Solar Probe Cup (SPC) \citep{Case:2020} with 3 top-hat electrostatic analyzers (ESAs) collectively called Solar Probe ANalyzers (SPAN), one that measures ions \citep[SPAN-I]{Livi:2021inrev} and 
two that measure electrons (\citep{Whittlesey:2020}, SPAN-e). While PSP's heat shield partially obscures the SPAN fields-of-view (FOVs), the joint SPAN-SPC FOVs cover the full sky.

This paper primarily uses SPAN-I 3D measurements of ion VDFs, resolved in $(E,\theta,\phi$) = (energy, elevation, azimuth) phase-space. The FOV limitations for SPAN-I are carefully considered when choosing appropriate time periods for optimal ion VDF analysis. For the observations described in this section, we use survey SPAN-I data at 7s cadence from PSP's fourth perihelion around the Sun (Encounter 4).
The electromagnetic wave activity is analyzed using the FIELDS instrument, using the highest resolution magnetometer data ($\approx$293 samples/s) sufficient to capture ion-scale dynamics in the narrow wave frequency band observed between 1 and 7 Hz in the electromagnetic fields data. Following the methodology in \cite{Torrence:1998,Bowen:2020a,Verniero:2020}, we extract spacecraft-frame wave power and polarization properties of the electromagnetic waves by applying a Morlet wavelet transform to the FIELDS magnetometer data.



\indent Encounter 4 revealed novel complexities in the SPAN-I measured VDFs compared to the ones observed in Encounter 2 (see Figure 2 of \cite{Verniero:2020}). \figref{fig:cbs}(a) shows an example of these new complex proton VDFs displaying the three core, beam, and hammerhead populations circled in orange, blue, and grey, respectively. The VDF was collapsed onto the $\theta$-plane in ($v_r$,$v_z$) SPAN-I instrument coordinates, where $v_r=v_x^2+v_y^2$. The black arrow represents the magnetic field direction rotated in the SPAN-I instrument frame, where the length is the \Alfven speed $v_A$ and the head is placed at the solar wind velocity. The extra asymmetry represented by the hammerhead population (which lies perpendicular to the magnetic field) was first noticed during a 7-hour period of quiescent solar wind accompanied by right-hand (RH) polarized ion-scale waves at $r=28 R_{\odot}$ in Encounter 4, as shown in \figref{fig:cbs}(b)-(e). Displayed is the FIELDS magnetometer observed (b) magnetic field $\V{B}$ in RTN coordinates, (c) wavelet spectrogram of $\V{B}$, and (d) perpendicular polarization (red = right-handed in the spacecraft frame, blue = left-handed). During this interval, the proton gyrofrequency was $\approx$1.6 Hz, so the narrowband wave power and polarization from \figref{fig:cbs}(c) and (d), as measured in the spacecraft frame, occur at the proton gyrofrequency scale. Doppler-shifting the wave frequencies from the spacecraft to the plasma frame is nontrivial in this interval since the solar wind Mach number was close to unity. Shown in \figref{fig:cbs}(e) is the SPAN-I measured temperature anisotropy $(T_\perp/T_\parallel)$ obtained by the plasma moments of the total proton VDF. Note that we do not obtain this value by diagonalizing the temperature tensor, because SPAN-I's obscured FOV and the resulting partial moments introduce systematic errors. Instead, $\V{B}$ is first rotated to the SPAN-I instrument frame and we then extract the perpendicular and parallel components of the temperature tensor with respect to $\V{B}$. Note that the presence of the large beam causes $(T_\perp/T_\parallel) < 1$ when only considering the moments of the ion VDF. As discussed in \cite{Klein:2021}, the fits of the proton VDFs hold more information on the nature of wave-particle energy transfer at kinetic scales.
\begin{figure}[t]
\hspace{.5in} (a) Density \hspace{3.0in} (b) Drift Velocity 
\vfill
{\includegraphics[trim = 0mm 0mm 0mm 0mm, clip, width=0.5\textwidth]{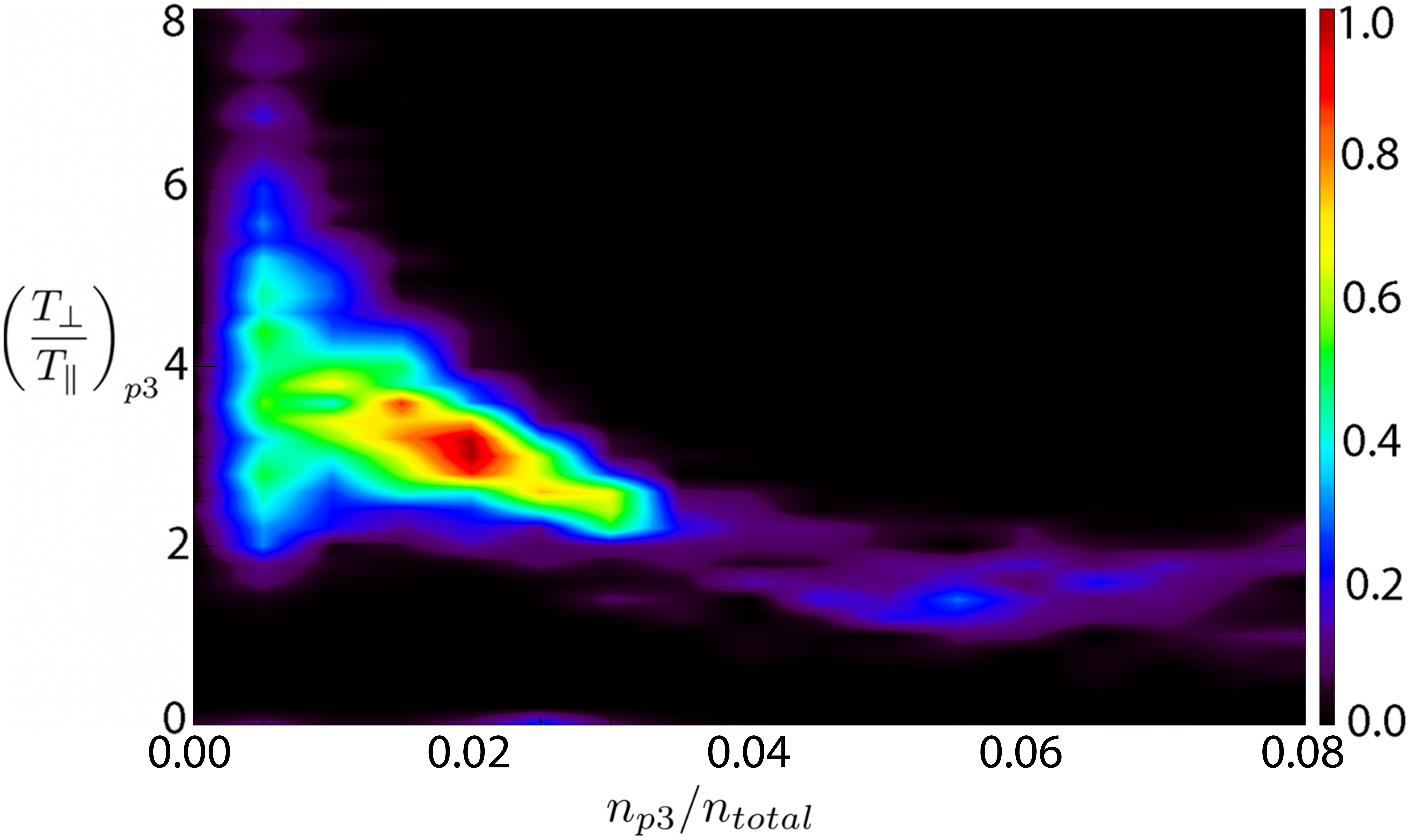}}
\hspace{.05in}
{\includegraphics[trim = 0mm 0mm 0mm 0mm, clip, width=0.5\textwidth]{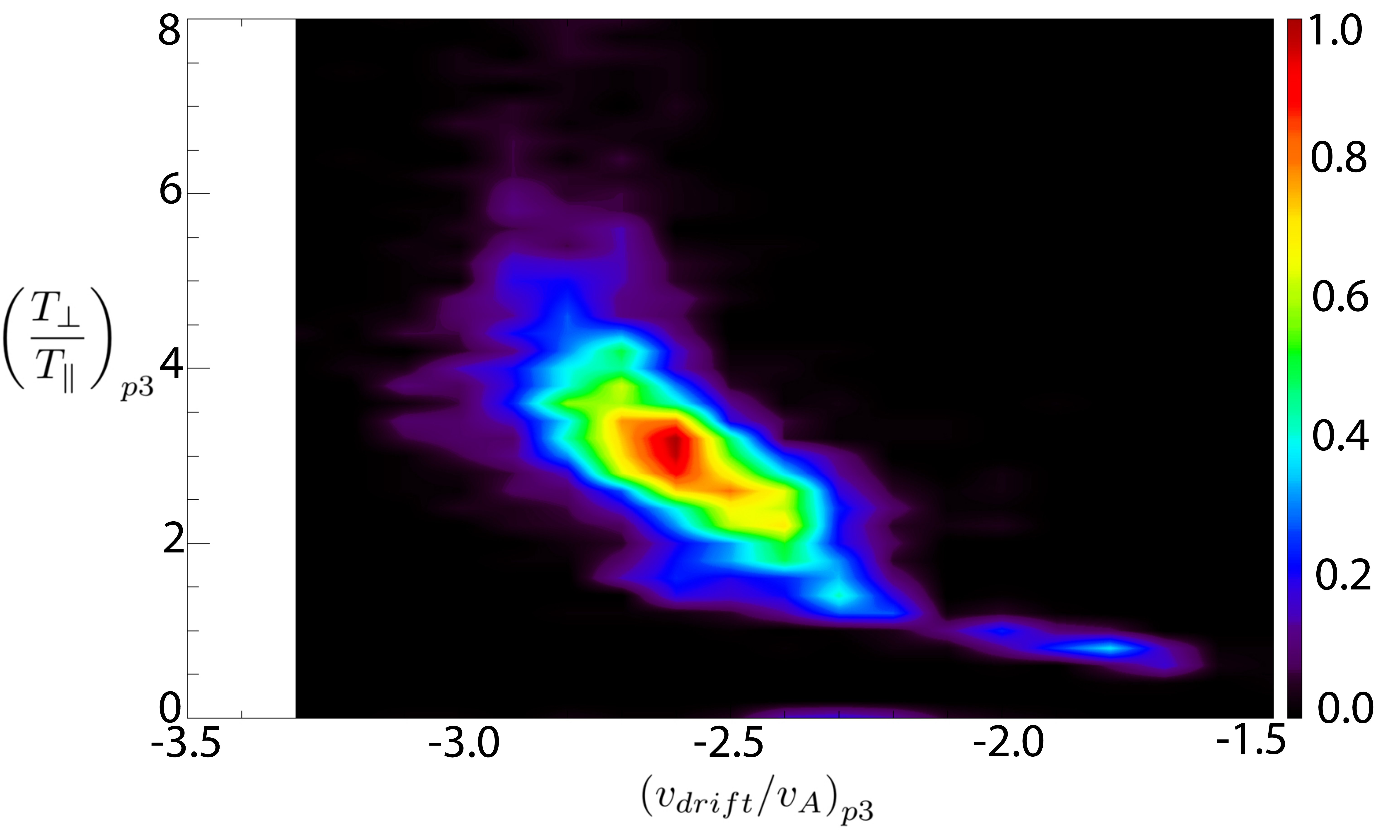}}
\caption{Initial characterization of the hammerhead population through trends of $(T_\perp/T_\parallel)_{p3}$ vs (a) fractional density and (b) relative drift velocity to the core population, normalized by the local \Alfven speed. Note that since the magnetic field vector points inwards in the SPAN-I instrument frame (see black arrow in \figref{fig:cbs}(a)), velocities have a negative sign during this time period. The color bar on each plot represents the number of fits, normalized by the total.
\label{fig:hist}}
\end{figure}

\indent Over the 7-hour period shown in \figref{fig:cbs}(b)-(e), we performed a triple bi-Maxwellian fit to the VDFs as a rudimentary approach to quantifying the observed tertiary population representing the proton particles participating in perpendicular phase-space diffusion, exhibiting a hammerhead feature. Standard Levenberg-Marquardt fits \citep{Levenberg:1944,Marquardt:1963} were applied to the L2 SPAN-I data; each portion of the proton VDF -- core (p1), beam (p2), and hammerhead (p3) -- is assigned to a bi-Maxwellian Gaussian function, $f_{pi}\vert_{(i=1,2,3)}$, that extracts the physical plasma parameters of density $n_{pi}$, velocity $\V{v}_{pi}$, and anisotropic temperature $(T_\perp,T_\parallel)_{pi}$. Similar to \citet{Alterman:2018}, the core population $f_{p1}$ is first fitted with a 3D velocity $(v_x,v_y,v_z)_{p1}$. The other two populations, $f_{p2}$ and $f_{p3}$, are assumed to be moving parallel to the magnetic field and fitted to a 1D drift velocity $(v_{drift})_{pi}$ relative to the core population. This procedure leads to 14 fitting parameters that can vary to obtain the best fit, up to physically reasonable constraints. First, the initial guess of each fitting parameter did not depend on the previous one, but rather was based on a reasonable fraction of the SPAN-I moments of density $\Tilde{n}$, velocity $\V{\Tilde{v}}$, and temperature $\Tilde{T}$. Next, the parameters $n_{pi}$ and $(T_\perp/T_\parallel)_{pi}$ for each population were constrained to be positive definite with not exceeding $3\Tilde{n}$ and $10\Tilde{T}$, respectively. Since the magnetic field points inwards in the SPAN-I instrument frame (see black arrow in \figref{fig:cbs}(a)), velocities have a negative sign, so we constrain $(v_{drift})_{pi}$ between $\pm 3\Tilde{v}_x$ (since $\Tilde{v_x}$ is nearly radial) . From \figref{fig:cbs}(a), we notice that the hammerhead population (grey) exhibits an oblong shape corresponding to a large temperature anisotropy, $(T_\perp/T_\parallel)_{p3}$. To ensure robust fit convergence, we additionally imposed a threshold $(T_\perp/T_\parallel)_{p1,2}$ of 6 for the core and beam populations and a threshold $(T_\perp/T_\parallel)_{p3}$ of 10 for the hammerhead population. If a given VDF fit exceeds these parameters, then it is discarded.

\figref{fig:hist} displays a preliminary occurrence rate quantification of the hammerhead population during the 7-hour ion-scale wave storm shown in \figref{fig:cbs}(b)-(e), and consequently a possible measure related to perpendicular velocity-space diffusion strength. We found that $(T_\perp/T_\parallel)_{p3}$ had a mean value of $\approx 2.5$. \figref{fig:hist} shows that the hammerhead population was (a) $\approx$ 2$\%$ of the total proton density and (b) drifting $\approx$ 2.5$v_A$ relative to the core population (a negative drift velocity is displayed since the magnetic field is pointing inwards in the SPAN-I instrument frame). From \figref{fig:hist}(a), we see that $(T_\perp/T_\parallel)_{p3}$ is asymptotic to 1 starting at $\approx 4 \%$. This suggests that there may be stronger perpendicular diffusion when the hammerhead density is less than 4\% of the total proton density; after this threshold, the hammerhead population relaxes toward isotropy and is indistinguishable from the beam population. This maximum density could also be a threshold for the hammerhead population to diffuse into other portions of phase-space.


\begin{figure}[t]
\centering
{\includegraphics[trim = 0mm 0mm 0mm 0mm, clip, width=0.85\textwidth]{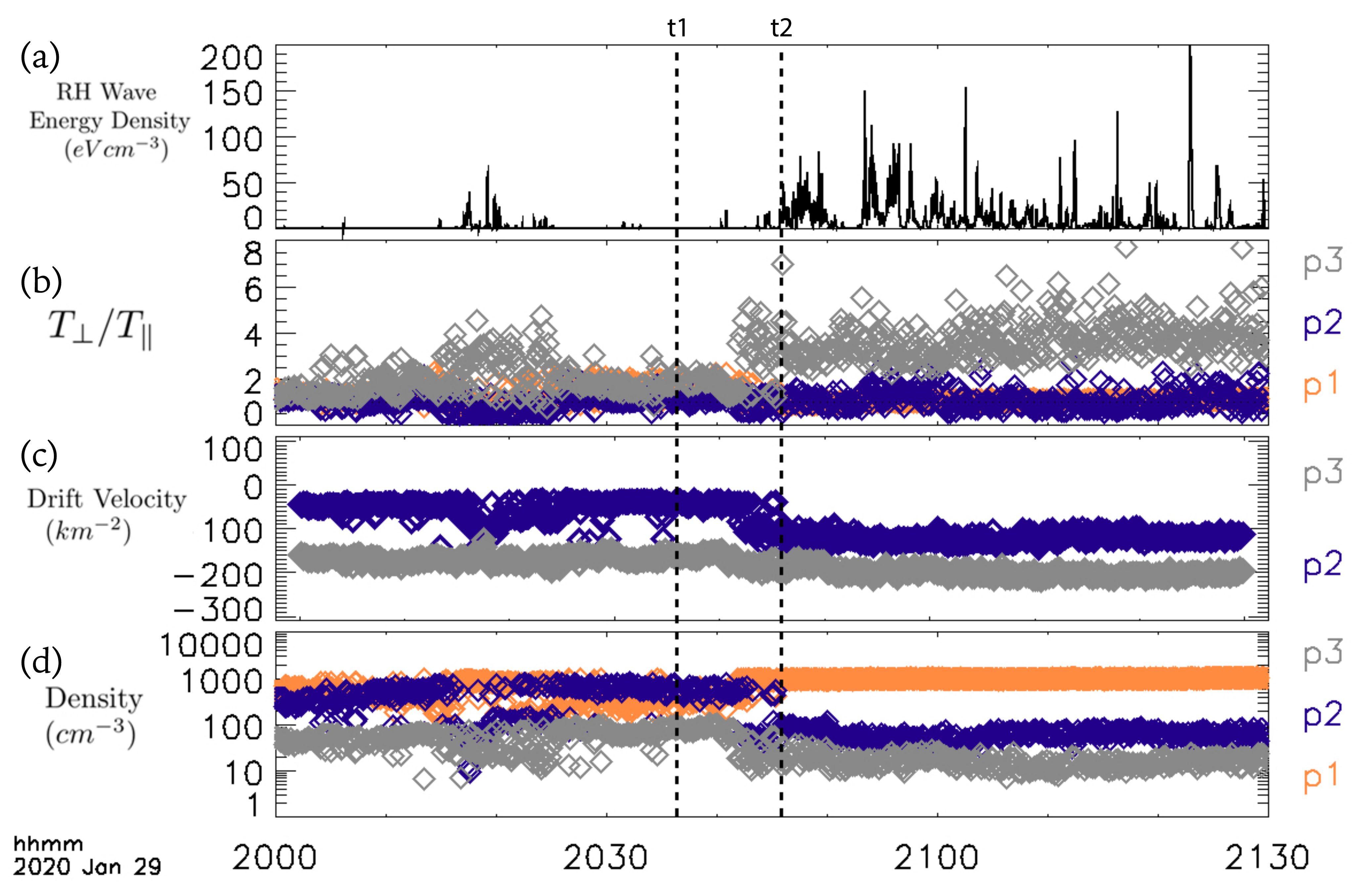}}
\caption{Period of intense wave enhancement showing (a) integrated Right-Handed Wave Power over 0.1-9 Hz, (2) temperature anisotropy fits of beam (blue) and hammerhead (grey), (3) drift velocity fits of beam (blue) and hammerhead (grey) with respect to the core, (4) density fits of core (orange), beam (blue), and hammerhead (grey).
\label{fig:rhwv}}
\end{figure}

\begin{figure}
\hspace{1in} (a) Proton VDF at t1 \hspace{1.9in} (b) Proton VDF at t2
\vfill
\centering
{\includegraphics[trim = 0mm 0mm 0mm 0mm, clip, width=0.36\textwidth]{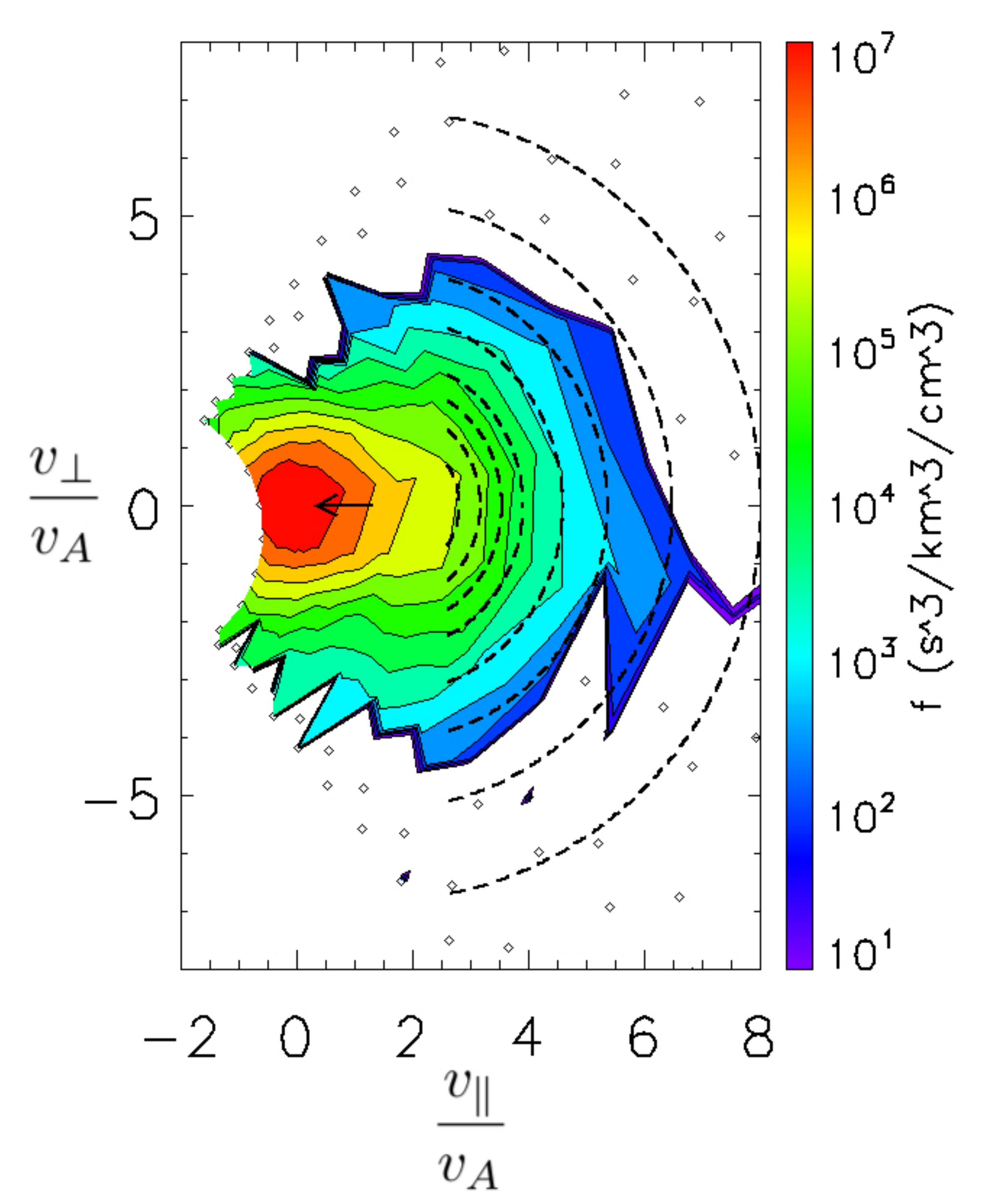}}
\hspace{.5in}
{\includegraphics[trim = 0mm 0mm 0mm 0mm, clip, width=0.36\textwidth]{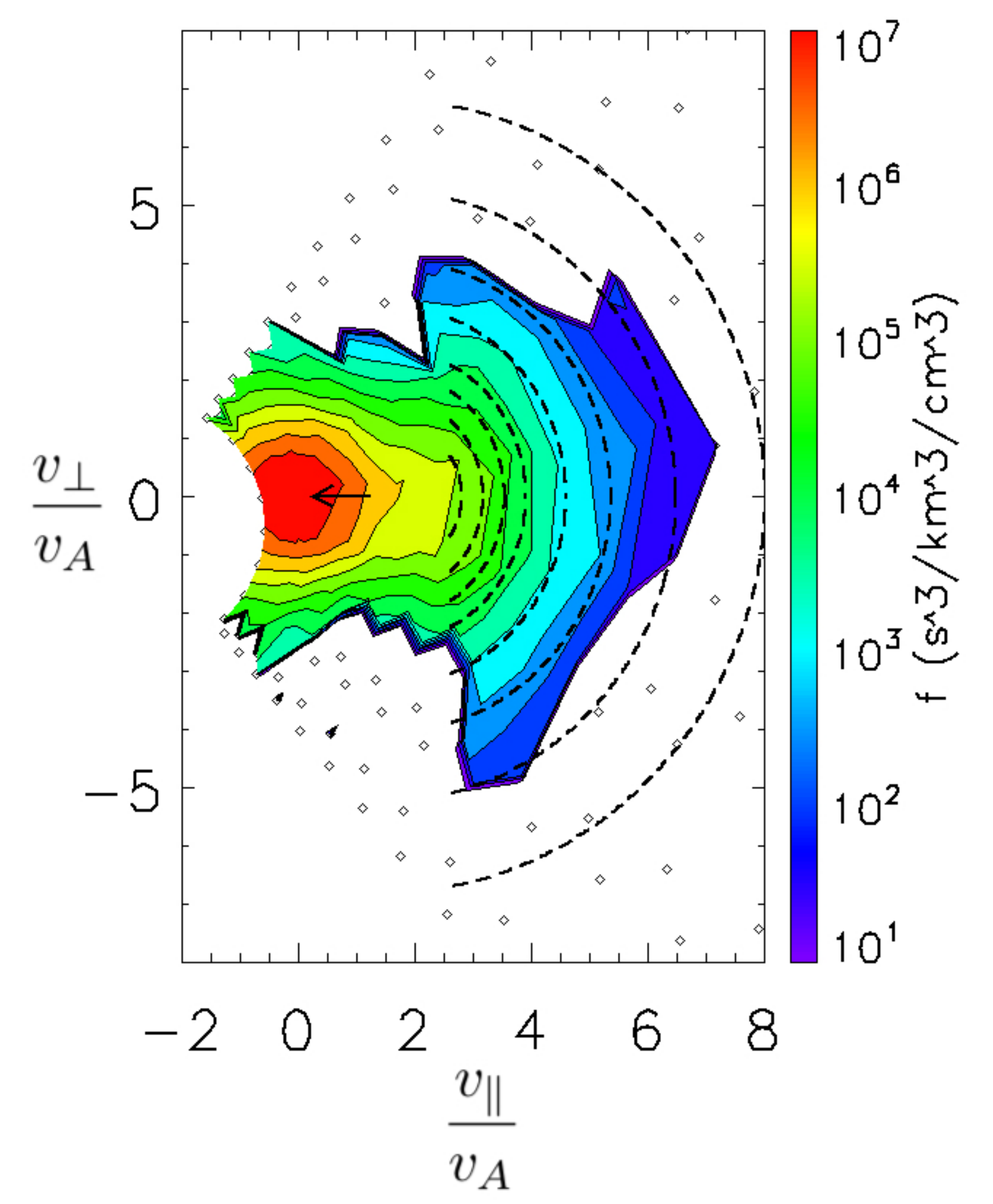}}
\caption{Perpendicular velocity-space diffusion in proton VDFs during period of intense wave enhancement shown in \figref{fig:rhwv}. Panels (a) and (b) shows the VDFs in magnetic field-aligned coordinates at $t1$ and $t2$, respectively, indicated by the dashed black lines in \figref{fig:rhwv}. The black arrow in each panel indicates the direction of the magnetic field, rotated to SPAN-I instrument coordinates; the head of the arrow is placed at the solar wind velocity, with length representing the local \Alfven speed. 
\label{fig:rhwv_vdf}}
\end{figure}

During the 7-hour ion-scale wave storm period displayed in \figref{fig:cbs}(b)-(e), the most intense enhancement of ion-scale wave power activity occurred during 2020-01-29/20:40:00-21:30:00. \figref{fig:rhwv}(a) displays the integrated RH wave power over the ion-scale frequency range of 0.1-9 Hz during this time period. Panels \figref{fig:rhwv}(b)-(d) show the 3-component fits of temperature anisotropy, drift velocity, and density, respectively. One can notice from \figref{fig:rhwv}(b) that at around 2020-01-29/20:20:00, the value of $(T_\perp/T_\parallel)_{p3}$ doubled from $\approx$1.5 to 3. This is also concurrent with the period of enhanced integrated RH wave power in \figref{fig:rhwv}(a). At time $t1$, indicated by the vertical black dashed black line, the wave activity decreased, and $(T_\perp/T_\parallel)_{p3}$ decreased back to $\approx$1.3. As shown by \figref{fig:rhwv_vdf}(a), the proton VDF does not display a hammerhead feature at $t1$, and the VDF appears more isotropized. On the other hand, once the RH wave activity enhances at $t2$, we see from \figref{fig:rhwv_vdf}(b) that the hammerhead feature in the VDF is more pronounced and follows the black dashed contour arcs. The nature of these predicted curves are explained in \secref{sec:theory}. During the enhanced ion-scale wave period of 2020-01-29/20:40:00-21:30:00, \figref{fig:rhwv}(b) shows that $(T_\perp/T_\parallel)_{p3}$ increased again, with a mean value of $\approx$3.7, and fluctuations up to $\approx$8. Thus, these preliminary fits suggest that enhanced ion-scale RH circularly polarized wave activity may be associated with the occurrence rate of the hammerhead feature in the proton VDFs.

\figref{fig:rhwv}(c) also shows that the drift velocity of the hammerhead population (relative to the core) increased by ~$0.5v_A$ from before $t1$ and after $t2$. The observed increase in drift velocity for both the beam and hammerhead populations after $t2$ demonstrates that the overall proton strahl drift is coincident with the enhanced RH ion-scale wave power. One also notices from \figref{fig:rhwv}(d) that after $t2$, the density of both the beam (blue) and hammerhead (orange) population decreased, while the core (red) density increased. This may be a consequence of the proton VDF fitting procedure. For example, before $t2$, when the temperature anisotropy is small, the distinction between $p2$ and $p3$ is less significant. Assuming both $p2$ and $p3$ have substantial drift velocities, both $p2$ and $p3$ can be considered part of the `beam', as seen in \figref{fig:rhwv_vdf}(a). As a consequence, a large temperature anisotropy for $p3$ is thus a necessary condition in order diagnose the presence of a hammerhead feature. Note that the preceding argument may also explain the density cutoff from \figref{fig:hist}(a). Within the subset of the data with large $(T_\perp/T_\parallel)_{p3}$, the density of $p3$ could determine the strength of the hammerhead, and subsequently a more quantitative measure of the velocity-space diffusion signature. But, for a binary diagnostic for the hammerhead occurrence rate, we find that $(T_\perp/T_\parallel)_{p3}$ is optimal.

As more evidence that these hammerhead features are concurrent with the RH ion-scale wave storm enhancement, we note that in the period 2020-01-29/16:00:00-23:00:00 where the hammerheads are seen, the peak wave power appears at $\approx$4.198 Hz. At and around this frequency, the degree of circular polarization was $\approx$0.722. The degree of circular polarization at 4.198 Hz frequency computed over the whole day was $\approx$0.056, yielding a ratio of $\approx$12.8. Thus, these complex asymmetries in the SPAN-I proton VDFs suggest they are associated with RH ion-scale waves.

Motivated by the enhanced levels of right-circularly polarized, ion-scale waves during this hammerhead event, we explore in the next section a theoretical model for the origin of hammerhead features that relies on resonant interactions between beam ions and right circularly polarized ion-scale waves.

\section{Theory: How Resonant Interactions Between Protons and Fast-Magnetosonic/Whistler Waves Generate Hammerhead Features in Proton VDFs}
\label{sec:theory}

As mentioned in the introduction, particle beams provide a source of free energy that can cause different plasma waves to become unstable. The resonant interaction between the growing waves and the beam particles cause the particles to diffuse in velocity space. This diffusion drives the particle distribution toward a marginally stable state; afterward, the wave growth halts. The iso-surfaces in velocity space along which the particle distribution function~$f$ is constant in this marginally stable state are referred to as `kinetic shells' \citep{Isenberg:2001}. This relaxation toward a marginally stable state results in deceleration and perpendicular heating of the beam ions. The ions lose more energy through deceleration than they gain via perpendicular heating (see \figref{fig:ptcls_lose_energy}), and the net energy lost by the beam ions is transferred to the growing fast-magnetosonic/whistler (FM/W) waves. In this section, we explain in detail how this process works for ion beams interacting with parallel-propagating FM/W waves. A general solution to this problem was given by \cite{Isenberg:1996}, who presented detailed analytic results for the case in which the FM/W waves are described by the cold-plasma dispersion relation. We reproduce parts of their solution here and numerically compute the shape of the kinetic shells for the case in which the FM/W waves are described by the hot-plasma dispersion relation, which is more relevant to inner heliospheric conditions sampled by PSP.

As alluded to in the previous paragraph, we confine our discussion to the `slab-symmetric' case in which the wave vector is parallel to the background magnetic field $\bm{B}_0$, assumed in the $z$ direction:
\begin{equation}
    \bm{B}_0 = B_0 \bm{\hat{z}}.
\end{equation}
Although we consider the interaction of FM/W waves with proton beams, we neglect these beams as well as all heavy ions when determining the dispersion relation of the waves. For the mathematical expressions that follow, we work in the reference frame in which the proton bulk velocity vanishes. We write the FM/W dispersion relation in the form
\begin{equation}
    y = F(x),
    \label{eq:disp_general}
\end{equation}
where
\begin{equation}
    y = \omega/\Omega_{\rm p} \qquad x = k_z v_{\rm A}/\Omega_{\rm p},
\end{equation}
$\omega$ is the wave frequency, $k_z$ is the component of the wave vector along $\bm{B}_0$ (which is the only non-vanishing wave-vector component in our analysis), $\Omega_{\rm p}=q B/m_{\rm p} $ is the proton cyclotron frequency, and $v_{\rm A}=B/\sqrt{4 pi n_{\rm p} m_{\rm p}}$ is the Alfv\'en speed. 
In the cold-plasma approximation for a proton-electron plasma, $F(x)$ is given by
\begin{equation}
    F(x) = \frac{x^2}{2} \left( 1 + \sqrt{1 + \frac{4}{x^2}}\right) \hspace{0.5cm} \mbox{(cold plasma)},
    \label{eq:disp}
\end{equation}
where $\omega$ is assumed much less than the electron cyclotron frequency.

Waves and ions interact resonantly only when they satisfy the resonance condition
\begin{equation}
    \omega - k_z v_z = n \Omega,
    \label{eq:res}
\end{equation}
where $v_z$ is the component of the particle velocity along $\bm{B}_0$, $\Omega$ is the ion cyclotron frequency, and $n$ is any integer. For the present right-handed and parallel FM/W waves under consideration, $n=-1$. The left-hand side of Equation~(\ref{eq:res}) is the wave frequency Doppler\blue{-}shifted into the reference frame that moves along $\bm{B}_0$ at the same velocity as the ion, which we call the `parallel co-moving frame.'

For concreteness, we take $\omega$ to be positive. (This entails no loss of generality; negative frequencies can be accommodated by flipping the sign of $k_z$ in the discussion to follow.)  We assume that the beam ions satisfy $v_z > 0$. (Again this implies no loss of generality.) We then restrict our attention to FM/W waves traveling in the $+z$ direction, i.e., with positive values of $\omega/k_z$, which amounts to taking $k_z >0$, given our assumption that $\omega > 0$. Interactions between FM/W waves and beam ions propagating in the opposite direction can occur, but they generally give rise to particle energization and hence wave damping, and are thus not relevant to the scenario we are exploring in this section, in which resonant wave-particle interactions amplify FM/W waves while decelerating beam ions \citep[see][ for a further discussion]{Verscharen:2013c}.

Parallel-propagating, right circularly polarized FM/W waves with $\omega>0$ resonate with ions only via the $n=-1$ resonance in Equation~(\ref{eq:res}) \citep{Stix:1992, Tsurutani:1997}. To understand the importance of the $n=-1$ resonance for ion-FM/W interactions, it is useful to consider a 3D coordinate system $(x,y,z)$ transformed to $(E_x(z),E_y(z),z)$-space, where electric field components $E_x$ and $E_y$ are evaluated as functions of~$z$ along some particular field line of~$\bm{B}_0$. This resulting curve is a helix centered on the $z$ axis, which we refer to as the `electric-field corkscrew.' The tilt of this corkscrew is determined by the requirement that, as the corkscrew travels past the $z=0$ plane in the $+z$ direction, the electric field vector in the $z=0$ plane undergoes right circularly polarized rotation. Because ions undergo left circularly polarized Larmor rotation, an ion that remains in the $z=0$ plane (i.e., at $z=0$ with $v_z=0$) does not resonate with an FM/W wave as it encounters a time-varying electric field. On the other hand, if $v_z > \omega/k_z$ for some ion, then in the `parallel co-moving frame' for that ion (which moves at velocity~$v_z$ along~$\bm{B}_0$), the electric-field corkscrew appears to propagate in the $-z$ direction. If the ion then fulfils $\omega-k_zv_z=-\Omega$, in the parallel co-moving frame, the electric field vector appears constant, which enables the FM/W wave to resonate with the ion. When $n=-1$, the Doppler-shifted frequency $\omega - k_z v_z$ is negative, implying that $v_z > \omega/k_z$. 

It follows from the foregoing discussion that we set
\begin{equation}
    \Omega = b \Omega_{\rm p}
    \label{eq:defb}
\end{equation}
in Equation~(\ref{eq:res}), where $b=1$ corresponds to proton beams. We then re-write Equation~(\ref{eq:res}) as
\begin{equation}
    y = -b + xu,
    \label{eq:res2}
\end{equation}
where
\begin{equation}
    u = \frac{v_z}{v_{\rm A}}.
    \label{eq:defu}
\end{equation}
 
\begin{figure}
 \hspace{.45in} (a)  \hspace{3.4in} (b) 
\vfill
{\includegraphics[trim = 0mm 0mm 0mm 0mm, clip, width=0.5\textwidth]{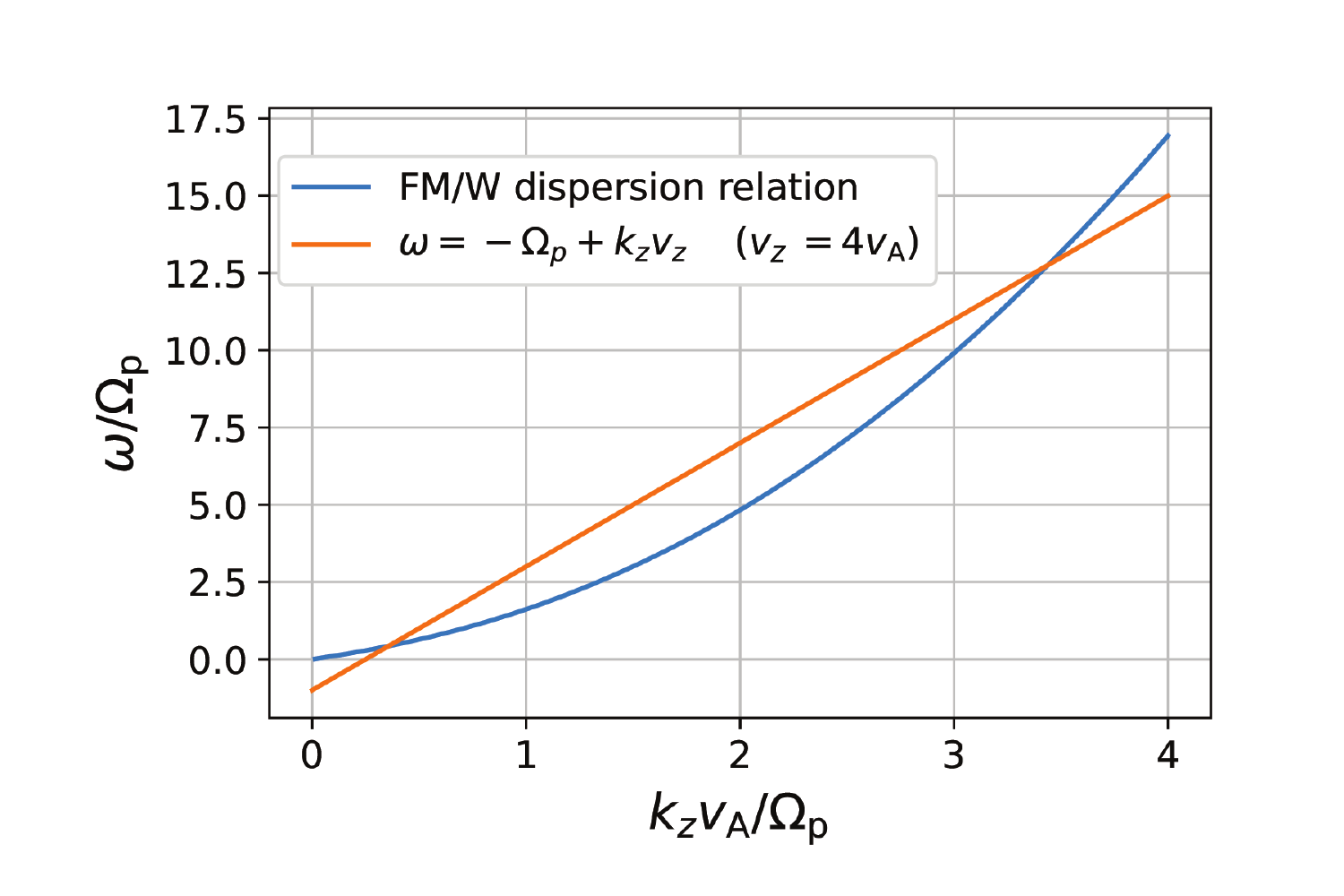}}
\hspace{.05in}
{\includegraphics[trim = 0mm 0mm 0mm 0mm, clip, width=0.5\textwidth]{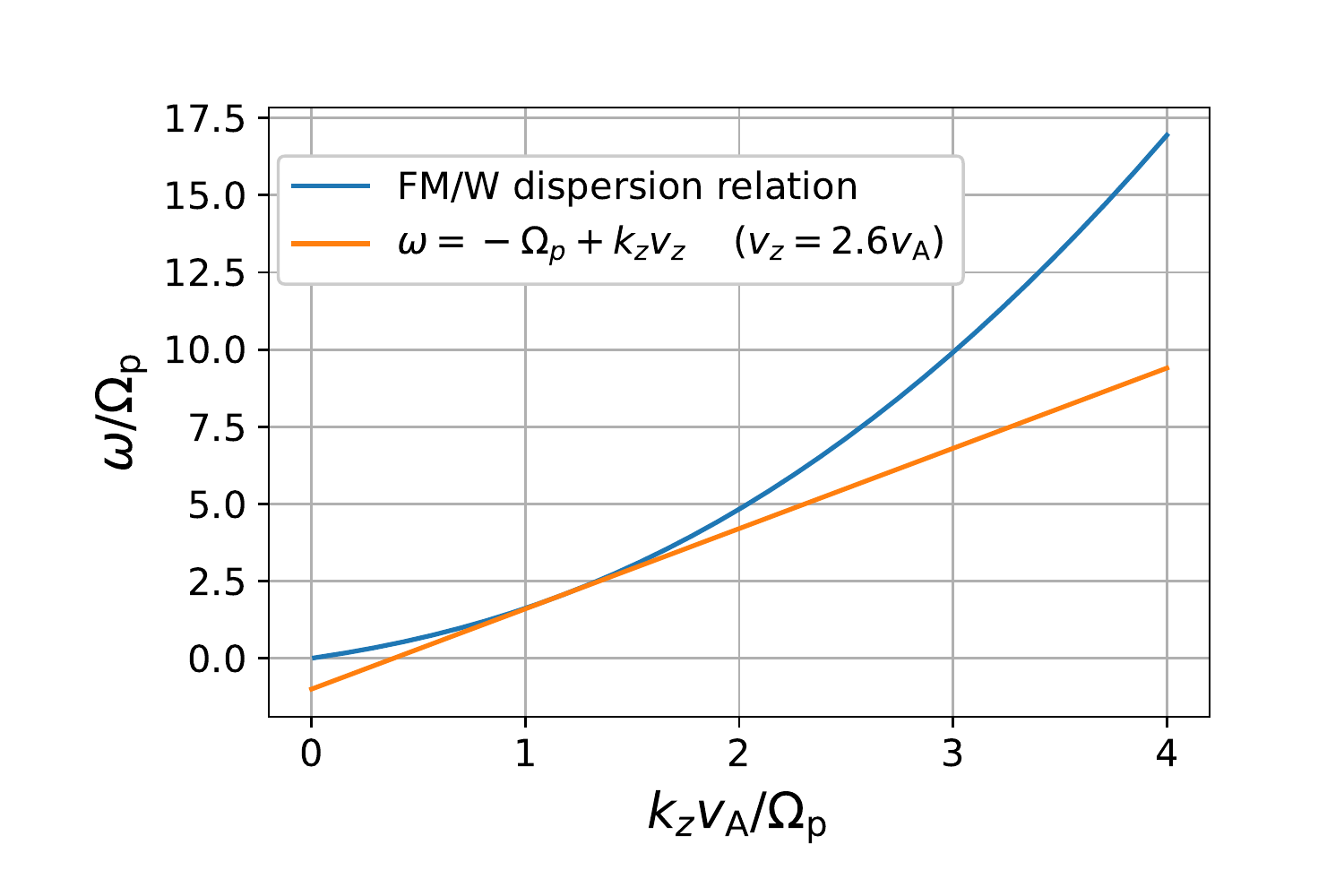}}
\caption{The cold-plasma FM/W dispersion relation (Equation~(\ref{eq:disp})) and the wave-particle resonance condition for a proton beam ($b=1$ in Equation~(\ref{eq:res2})). Resonant wave--particle interactions can occur when both curves intersect. Panel~(a) corresponds to protons with parallel velocity~$v_z = 4 v_{\rm A}$. Panel~(b) corresponds to protons with $v_z = 2.6 v_{\rm A}$, which is the minimum value of~$v_z$ for which protons can resonate with FM/W waves in the cold proton-electron-plasma approximation.
\label{fig:resonance_conditions}}
\end{figure}

The combinations of $(x,y)$ that fulfill both Equation~(\ref{eq:res2}) and Equation~(\ref{eq:disp_general}) gives the wave numbers and frequencies at which FM/W waves resonantly interact with ions with parallel velocity~$v_z$.
Panel~(a) of Figure~\ref{fig:resonance_conditions} illustrates this condition  for beam protons ($b=1$ in Equation~(\ref{eq:res2})) with $v_z = 4 v_{\rm A}$ and FM/W waves described by the cold-plasma dispersion relation.  
If we were to decrease~$v_z$, then the slope of the resonance line (the solution of Equation~(\ref{eq:res2})) would decrease, leaving the $y$-intercept of this line unchanged. Below a certain value of~$v_z$, denoted by $v_{\rm min}$, the resonance line
no longer intersects the dispersion relation (as long as $0<\omega\ll \Omega_{\mathrm e}$). As a consequence, beam protons interact with slab FM/W waves only when $v_z \geq v_{\rm min}$. When the FM/W waves are described using the cold-plasma dispersion relation, $v_{\rm min} = 2.60 v_{\rm A}$. On the other hand, $v_{\rm min} = 2.68 v_{\rm A}$ when
$F(x)$ in Equation~(\ref{eq:disp}) is determined from the hot-plasma dispersion relation for a Maxwellian, single-temperature, proton-electron plasma with $\beta=1$ using the PLUME code \citep{Klein:2015a}, where $\beta$ is the ratio of plasma pressure to magnetic pressure. Panel~(b) of Figure~\ref{fig:resonance_conditions} reproduces panel~(a) but with~$v_z = 2.6 v_{\rm A}$.

When an ion resonates with an FM/W wave at a particular resonant wave number and frequency, the ion diffuses along a contour in the $v_\parallel-v_\perp$ plane that corresponds to constant ion kinetic energy in the reference frame that moves along $\bm{B}_0$ at the phase velocity $\omega/k_z$ of the wave \citep{Stix:1992}, the wave frame. Conservation of particle kinetic energy in the wave frame is a consequence of the fact that the electric and magnetic fields are constant in time in the wave frame. Faraday's Law then implies that $\boldsymbol{\nabla}\times \bm{E} = 0$, so that the electric field in the wave frame can be written as the gradient of a potential~$\Phi$. The relative velocity between the wave frame and proton frame is parallel to $\bm{B}_0$. As a consequence, transforming from the proton frame to the wave frame does not lead to the appearance of a constant electric field. Because an ion's total energy $m (v^\prime)^2/2 + q \Phi$ in the wave frame is constant (where $\bm{v}^\prime$ is the ion's velocity in the wave frame), $m (v ^\prime)^2/2$ is effectively conserved. (For a different explanation in terms of wave quanta and energy and momentum conservation, see \cite{Stix:1992}.)

\begin{figure}
{\includegraphics[trim = 0mm 0mm 0mm 0mm, clip, width=1.0\textwidth]{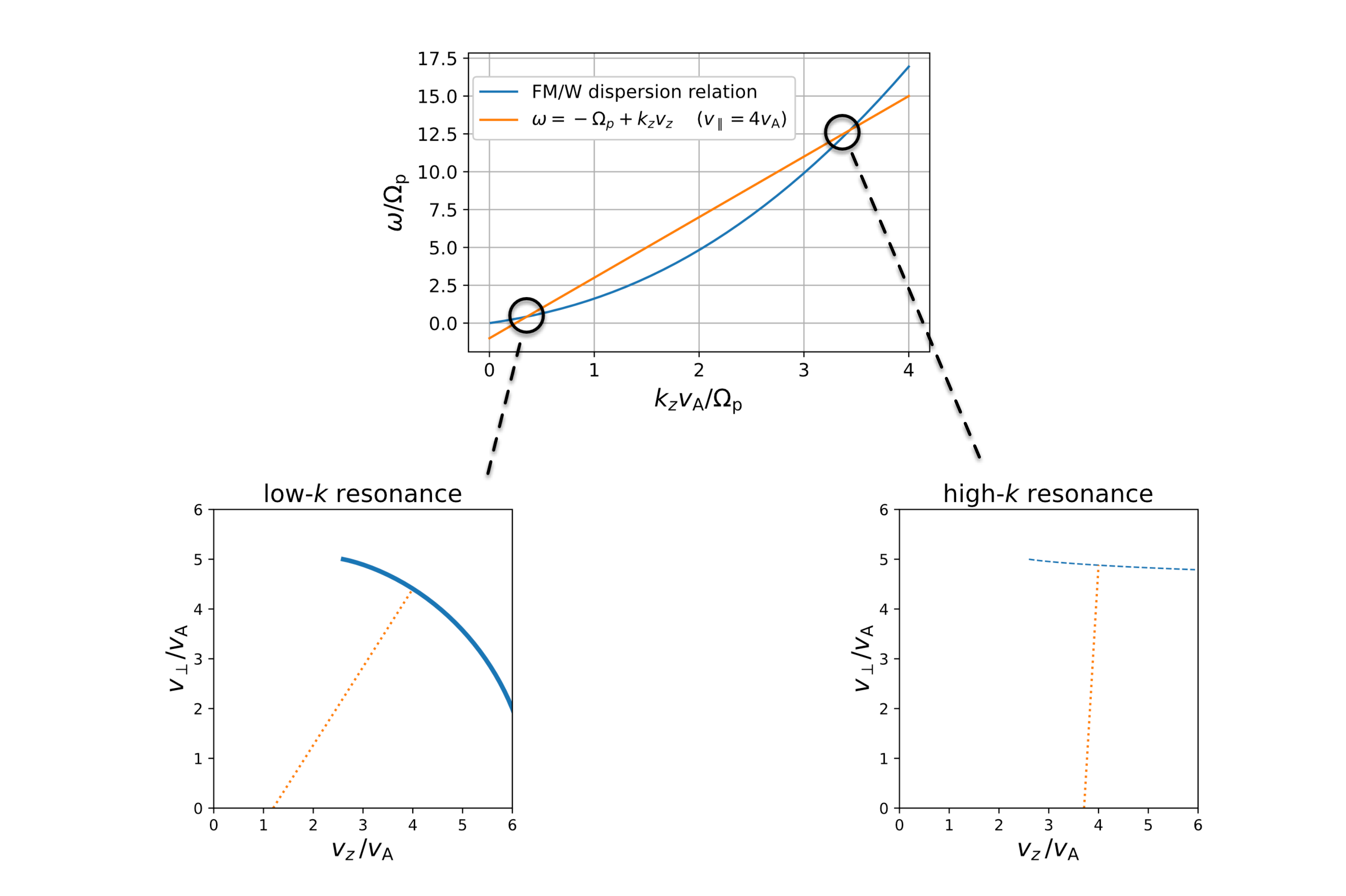}}
\caption{The top plot shows the FM/W dispersion relation (blue) intersecting with $n=-1$ resonance curve (orange). The lower left and lower right panels display the diffusion path cases for ions to resonate with FM/W waves via the low-k and high-k resonance, respectively, mapped to $(\frac{v_\perp}{v_A},\frac{v_z}{v_A})$-plane.
\label{fig:diffusion_contours}}
\end{figure}

Figure~\ref{fig:diffusion_contours} illustrates this energy-conserving velocity-space diffusion for the wave-particle resonances illustrated in \figref{fig:resonance_conditions}(a), corresponding to beam protons with $v_z = 4 v_{\rm A}$. As illustrated in this figure, beam protons with $v_z = 4 v_{\rm A}$ can resonantly interact with FM/W waves at two different wave numbers, one satisfying $k_z v_{\rm A} / \Omega_{\rm p} = 0.356$, and one satisfying $k_z v_{\rm A}/\Omega_{\rm p} = 3.44$, which we refer to as the low-$k$ resonance and high-$k$ resonance, respectively.
In the case of the low-$k$ resonance, the phase velocity $\omega/k_z$ of the resonant waves is $1.19 v_{\rm A}$.
{\em  If we imagine that protons and FM/W waves interact only via the low-$k$ resonance}, then
protons with $v_z = 4 v_{\rm A}$ diffuse along contours in the $v_\parallel-v_\perp$ plane that are locally tangent to semicircles centered at $v_\parallel = 1.19 v_{\rm A}$ and $v_\perp = 0$. This is illustrated by the solid-line diffusion contour in the lower-left plot of Figure~\ref{fig:diffusion_contours}. The entirety of the diffusion contour, however, is not semicircular, because the FM/W waves are dispersive. For example, as the particle's $v_z$ changes during the diffusion, so would the phase speed of the resonant waves, and thus the local tangent that describes the diffusion path.
The solid curve in the lower-left plot of Figure~\ref{fig:diffusion_contours} shows a diffusion contour that, at each $v_z$, is locally tangent to a semicircle centered at $v_\perp = 0$ and $v_z = v_{\rm phase}^{\rm (low)} (v_z)$, where $v_{\rm phase}^{\rm (low)}(v_z)$ is the value of $\omega/k_z$ for the FM/W waves that resonate with ions of a given~$v_z$ via the low-$k$ resonance.

For the high-$k$ resonance shown in Figure~\ref{fig:diffusion_contours}, $\omega/k_z$ is $3.71 v_{\rm A}$. 
{\em  If we imagine that ions and FM/W waves interact only via the high-$k$ resonance}, then an ion with $v_z = 4 v_{\rm A}$ interacting with FM/W waves via this high-$k$ resonance diffuses in the $v_\parallel - v_\perp$ plane along a contour that is locally tangent to a semicircle centered at $v_z = 3.71 v_{\rm A}$ and $v_\perp = 0$.  Like in the case of the low-k resonance, the phase speed of the resonant waves depends on $v_z$.
The dashed line in the lower-right plot in \figref{fig:diffusion_contours} shows a diffusion contour that, at each $v_z$, is locally tangent to a semicircle centered at $v_\perp = 0$ and $v_z = v_{\rm phase}^{\rm (high)}(v_z)$, where $v_{\rm phase}^{\rm (high)}(v_z)$ is the value of $\omega/k_z$ for the FM/W waves that resonate with ions of a given~$v_z$ via the high-$k$ resonance.

\begin{figure}
 \hspace{.45in} (a)  \hspace{3.4in} (b) 
\vfill
{\includegraphics[trim = 0mm 0mm 0mm 0mm, clip, width=0.5\textwidth]{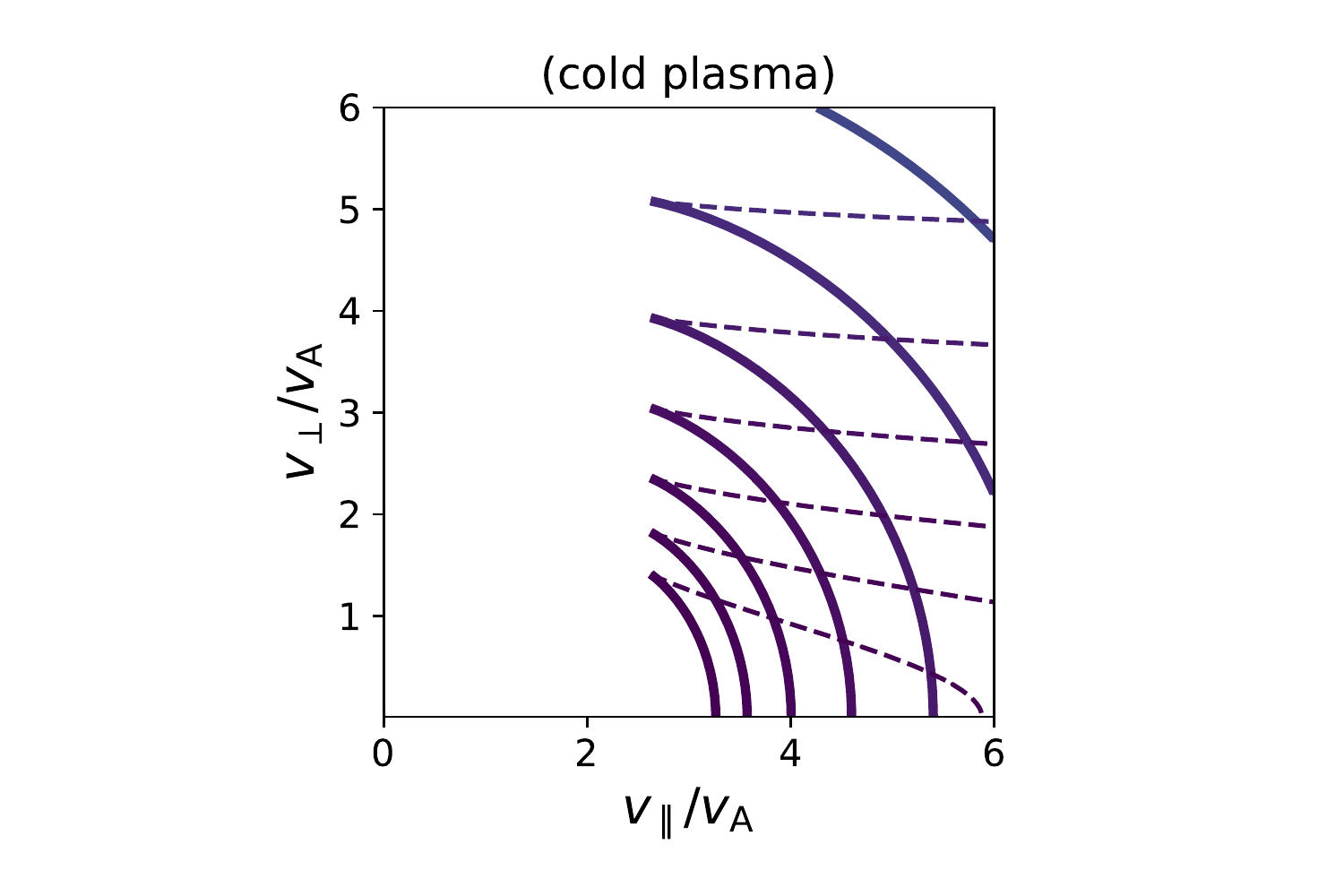}}
\hspace{.05in}
{\includegraphics[trim = 0mm 0mm 0mm 0mm, clip, width=0.5\textwidth]{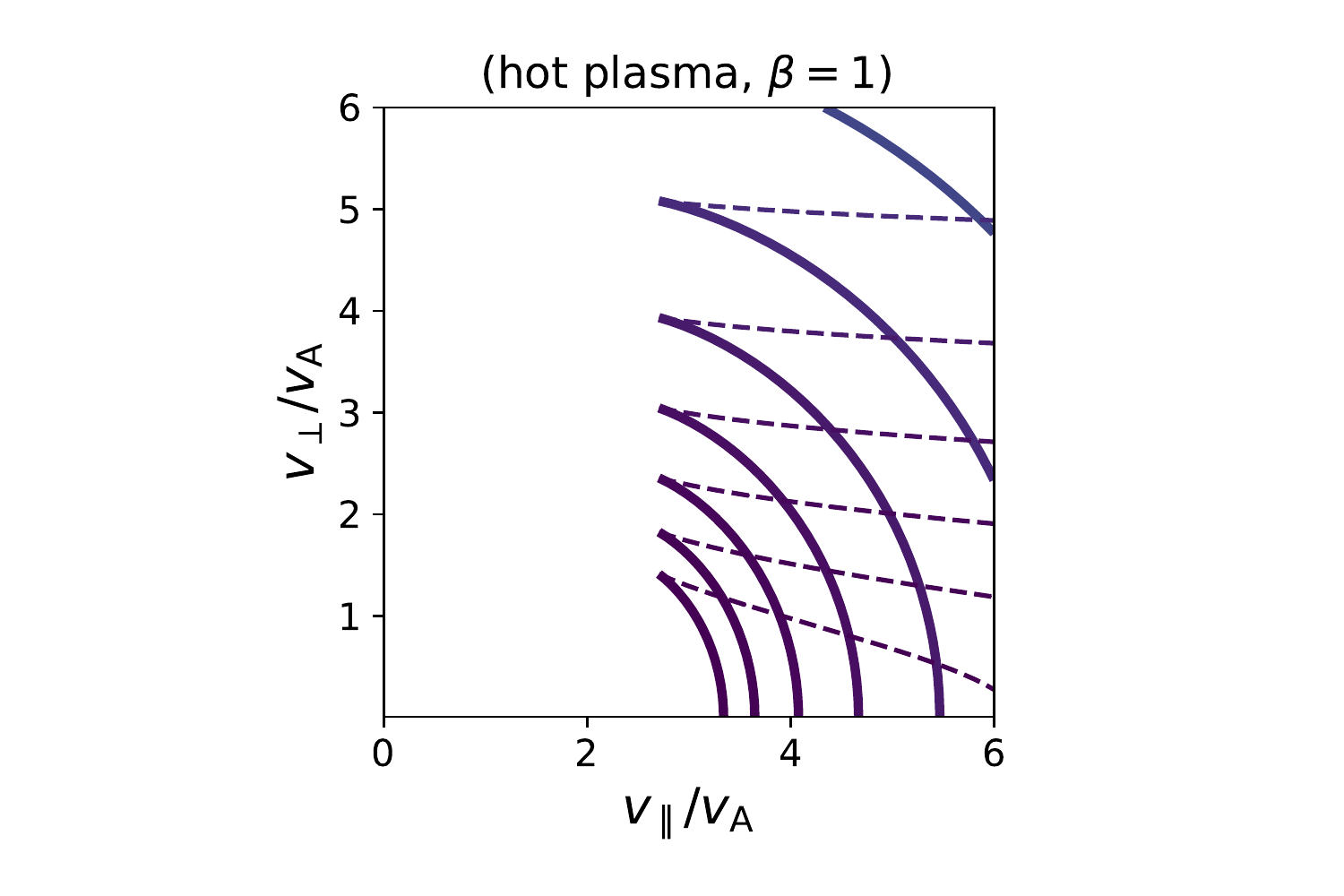}}
\caption{Velocity-space diffusion contours corresponding to the low-$k$ resonance (solid lines) and high-$k$ resonance (dashed lines). In panel~(a), the FM/W waves obey the cold-plasma dispersion relation for a proton-electron plasma (Equation~(\ref{eq:disp})). In panel~(b), the  
FM/W waves obey the hot-plasma dispersion relation for a Maxwellian, single-temperature, proton-electron plasma with $\beta=1$. The direction of the diffusion along these contours is determined by the condition that the diffusion occurs from larger to smaller values of $f(v_z,v_\perp)$.
\label{fig:kinetic_shells}}
\end{figure}

\begin{figure}
\hspace{.45in} (a)  \hspace{3.4in} (b) 
\vfill
\raisebox{1in}{\includegraphics[trim = 0mm 0mm 0mm 0mm, clip, height=2.5in]{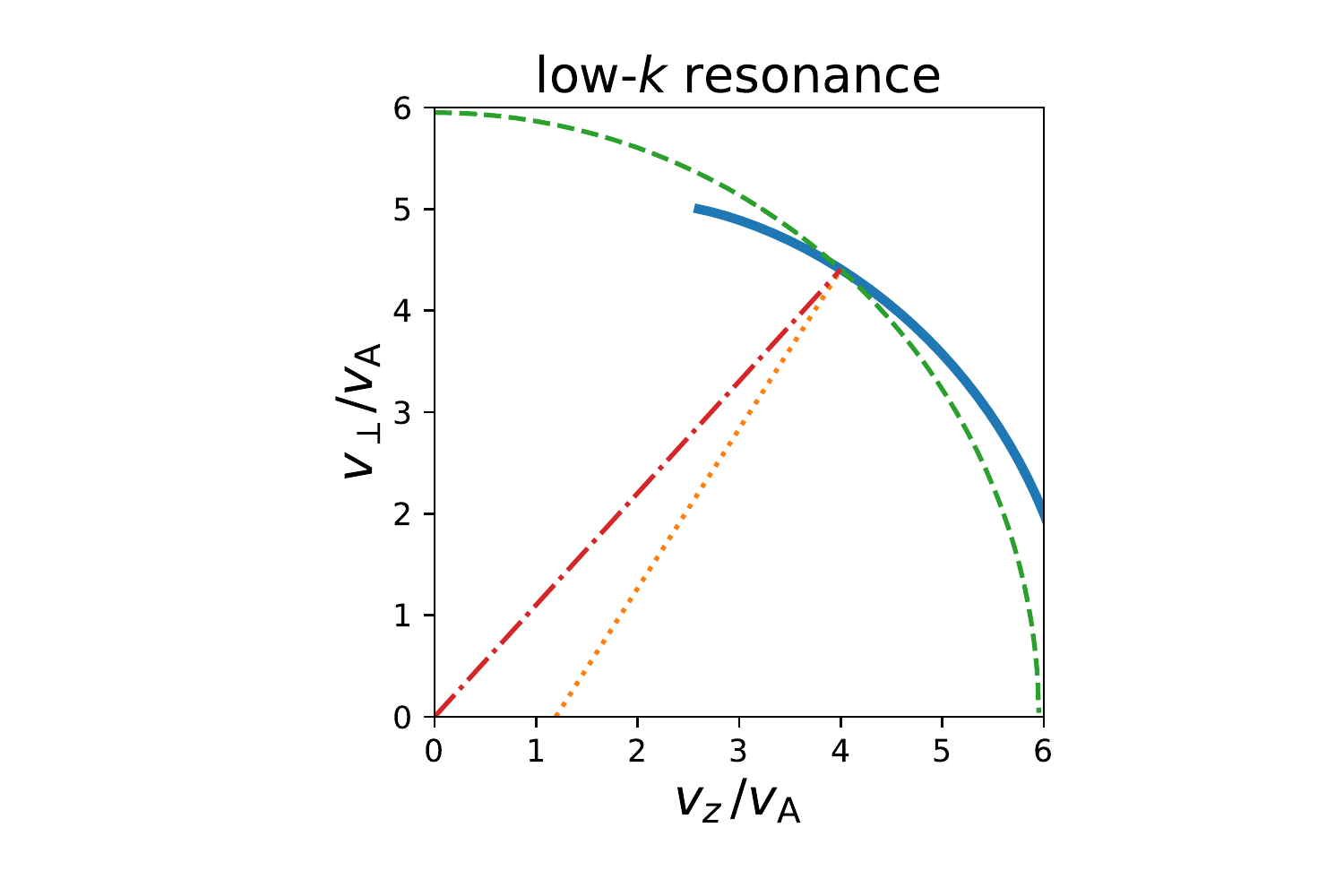}}
\hspace{0.05in}
{\includegraphics[trim = 0mm 0mm 0mm 0mm, clip, height=4in]{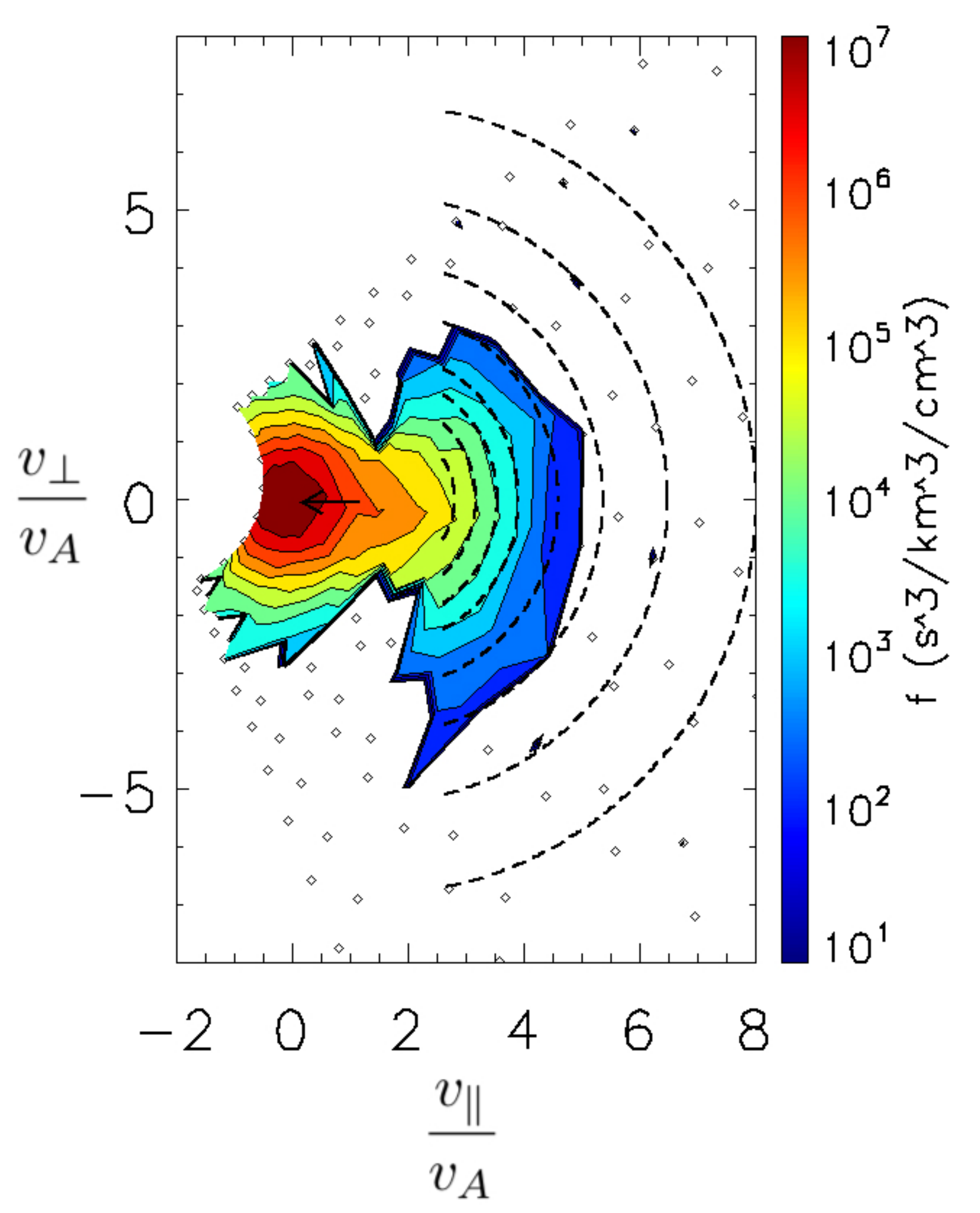}}
\caption{{\em Panel (a):} The solid line represents the same diffusion contour for the low-$k$ resonance that was pictured in the lower-left plot of Figure~\ref{fig:diffusion_contours}, which is locally tangent to a semicircle centered on the velocity coordinates of FM/W waves that resonate with the ions at each different~$v_z$. The dashed line is a semicircle of constant energy in the plasma rest frame, which is centered on the origin. As particles diffuse upward along the diffusion contour, their total kinetic energy decreases. {\em Panel (b):} example proton VDF in field-aligned plasma-frame coordinates, over-plotted with dashed the low-$k$ resonance contours when the FM/W waves are described using the cold-plasma approximation.
\label{fig:ptcls_lose_energy}}
\end{figure}

If, as in the preceding paragraphs, we suppose that only one of the two resonances (either low-$k$ or high-$k$) is active, then
the velocity-space contours along which the particles diffuse $v_\perp(v_z)$ can be computed as follows.
 The line that connects an ion's velocity-space coordinates $(v_z, v_\perp)$ to the velocity-space coordinates $(v_{\rm phase}(v_z),0)$ of the wave frame 
 of the waves that resonate with that ion has a slope of $v_\perp/[v_z - v_{\rm phase}(v_z))]$, where $v_{\rm phase}(v_z)$ is either $v_{\rm phase}^{\rm (low)}$ or $v_{\rm phase}^{\rm (high)}$, depending on which of the two resonances is assumed to be active. This line is perpendicular to the diffusion contour described by the function $v_\perp(v_z)$, from which it follows that 
\begin{equation}
    \frac{dv_\perp}{dv_z} = \frac{v_{\rm phase}(v_z) - v_z}{v_\perp}.
    \label{eq:slope2}
\end{equation}
(This point is nicely illustrated in Figure~2 of \cite{Isenberg:1996}.) Equation~(\ref{eq:slope2}) can be integrated to yield
\begin{equation}
 \frac{v_\perp^2}{2} = - \frac{v_z^2}{2} + \int v_{\rm phase} (v_z) {\rm d}v_z + C.
 \label{eq:kinetic_shell_eq}
\end{equation}
where $C$ is an integration constant representing an arbitrary offset (for simplicity, we set $C$=0). \cite{Isenberg:1996} evaluated Equation~(\ref{eq:kinetic_shell_eq}) analytically for the low-$k$ resonance using the cold-plasma dispersion relation in Equation~(\ref{eq:disp}), obtaining a parametric solution for $v_\perp(v_z)$ given in their Equations (14) and (15). Some examples of the resulting contours are shown in \figref{fig:kinetic_shells}(a), which also includes (numerical) solutions for the diffusion contours associated with the high-$k$ resonance. \figref{fig:kinetic_shells}(b) shows the diffusion contours that result when $F(x)$ in Equation~(\ref{eq:disp_general}) is determined from numerical solutions to the hot-plasma dispersion relation for a Maxwellian, single-temperature, electron-proton plasma with $\beta=1$ using the PLUME code~\citep{Klein:2015a}.

If a population of beam ions initiates with a distribution function~$f(v_z, v_\perp)$ that extends to values of $v_z$ exceeding $2.6 v_{\rm A}$,
then the beam ions can resonantly interact with FM/W waves and diffuse along the contours pictured in \figref{fig:kinetic_shells}. Their diffusion direction (upward or downward) depends on the gradient of $f(v_z,v_\perp)$. The diffusion would go upward (downward) if, when travelling upward (downward) along the diffusion path, the value of $f(v_z,v_\perp)$ decreases (increases). For concreteness, we assume the former condition is true for $f(v_z,v_\perp)$.

Because the ions diffuse from regions of higher particle concentration to regions of lower particle concentration in velocity space, they will diffuse upward along the low-$k$ resonance contours pictured in \figref{fig:kinetic_shells} towards larger values of~$v_\perp$. Although these ions experience an increase in $v_\perp$, they undergo parallel deceleration, and an ion's total kinetic energy decreases as it diffuses up along one of the low-$k$ resonance contours. This net loss of energy is illustrated in panel~(a) of \figref{fig:ptcls_lose_energy}. The energy lost by the resonant ions is transferred to the resonant FM/W waves, causing them to grow.
This process occurs on the quasilinear interaction timescale, which can be of order a few wave periods \citep{Isenberg:1996} and drives the particles towards a state in which the instability ceases to grow (i.e., the marginally stable state where diffusion contours are referred to as `kinetic shells' \citep[see, e.g.,][]{Isenberg:2001}. 

For the FM/W waves in the cold-plasma approximation, the kinetic shells end when $v_z$ drops to values $\simeq 2.6 v_{\rm A}$, below which the ions are unable to resonate with the FM/W waves. The level contours of such a marginally stable distribution have a `hammerhead' shape at $v_z \gtrsim 2.6 v_{\rm A}$, similar to some of the ion distribution functions seen by PSP, as illustrated in panel~(b) of  Figure~\ref{fig:ptcls_lose_energy}. This suggests that the hammerhead features in the observed VDFs are the result of proton-beam-driven instabilities of the parallel FM/W instability. For an initially Maxwellian beam, such an instability is only driven if the beam bulk speed in the proton bulk frame is greater than the parallel phase speed of the resonant wave \citep{Verscharen:2013a}.

In principle, the beam ions can also resonate with FM/W waves via the high-$k$ resonance. 
Due to the condition of diffusion toward smaller values of $f(v_z,v_\perp)$, the diffusing ions in that range would typically gain energy in the interaction. This kinetic energy would be drained from the resonant FM/W waves, causing those waves to damp. Because of this, we conjecture that the high-$k$ resonance is not significantly activated by the beam distributions encountered by PSP, which justifies our neglect of this second resonance when computing the kinetic shells.

\section{Conclusions and Future Work}
\label{sec:conc}

In this paper, we present new PSP observations of proton VDFs in which the proton beam is strongly broadened perpendicular to the magnetic field over a narrow range of parallel velocities, leading to VDF level contours with a hammerhead shape. In \secref{sec:obs}, we quantify this extra asymmetry by approximating the proton VDF as the sum of three components: a core, a narrow field-aligned beam, and a broadened `hammerhead', as seen in \figref{fig:cbs}(a). These new complex VDFs were first observed during a 7-hour right-hand circularly polarized wave storm at the ion gyrofrequency scale displayed in \figref{fig:cbs}(b)-(e). We explored how different properties of the hammerhead population (fractional density, temperature anisotropy, and drift Alfv\'en Mach number) correlate with one another during a 7-hour interval of PSP's fourth perihelion encounter with the Sun. These initial results from the triple bi-Maxwellian SPAN-I VDF fitting during this event are shown in \figref{fig:hist}, which suggest that the temperature anisotropy of the hammerhead population is the most consequential parameter. Future studies will use this diagnostic to produce a statistical characterization of the hammerhead occurrence rate as a function of various solar wind parameters, yielding insight on the nature of ion velocity-space diffusion as a solar wind heating mechanism.

The prevalence of the hammerhead features in the proton VDF is also coincident with the energy density of ion-scale waves that are right circularly polarized in the spacecraft frame. This is seen by observing the association of the enhancement of RH wave energy density in \figref{fig:rhwv}(a) with the change in proton VDF fitting parameters displayed in \figref{fig:rhwv}(b)-(d). It was again found that the increase in temperature anisotropy was most indicative of the presence of a hammerhead, as further illustrated by \figref{fig:rhwv_vdf}. Since the solar wind Mach number during the 7-hour wave event was close to unity, the approximation of Doppler-shifted frequencies from the spacecraft to the plasma frame is nontrivial (see \citep{Bowen:2020b}) and will be left for future work. A plasma dispersion analysis can be applied to the 3-component fits to better examine the components of the proton VDF leading to wave growth and/or decay. A more comprehensive wave analysis can also be performed to better characterize the correlation between the observed ion-strahl diffusion and right-hand circularly polarized waves over multiple encounters. The results presented in \secref{sec:obs} demonstrated a preliminary attempt at understanding the wave-particle energy transfer dynamics associated with the new asymmetries found in the proton VDFs.

In Section~\ref{sec:theory} we develop a theoretical model that explains the shape of these proton VDFs. This model is based on earlier work by \cite{Isenberg:1996} where beam protons cause right circularly polarized, parallel-propagating, FM/W waves to become unstable and grow. When a beam proton resonantly interacts with a FM/W wave, the proton's energy is conserved in the reference frame that propagates along the background magnetic field at the wave's phase velocity. This conservation law causes the protons to diffuse along a particular set of contours in the $v_z - v_\perp$ plane (where $v_z$ and $v_\perp$ are velocity components parallel and perpendicular to~$\bm{B}_0$) that can be computed based on the FM/W dispersion relation. Beam protons whose velocities are initially nearly aligned with~$\bm{B}_0$ diffuse along these contours to smaller values of $v_z $ and larger values of~$v_\perp$. The net effect of this parallel deceleration and perpendicular thermalization is to reduce the protons' kinetic energies (\figref{fig:ptcls_lose_energy}(a)). The energy lost by the protons is transferred to the FM/W waves, which causes the waves to grow. Velocity-space diffusion causes the proton distribution function~$f(v_z, v_\perp)$ to become constant along the aforementioned velocity-space diffusion contours, causing the plasma to become marginally stable to the growth of FM/W waves. The resulting constant-$f$ lines, or `kinetic shells' \citep{Isenberg:2001}, is consistent with the constant-$f$ lines in the hammerhead portion of the ion VDFs observed by PSP that are shown in Section~\ref{sec:obs}. This agreement, along with the association in the PSP observations between hammerhead features and right-circularly polarized ion-scale waves, suggest that the hammerhead VDFs seen by PSP indeed result from resonant interactions between beam protons and FM/W waves. We speculate that this process is more intense in the near-Sun environment, which is possibly why we do not observe the hammerhead features in VDFs measured at larger distances. The heliographic position of PSP, along with the SPAN-I field-of-view limitations will also be considered in future investigations.

The primary goal of this paper was to report the novel wave-particle PSP observations and to use linear theory to provide an intuitive explanation of their significance. Given the highly non-Maxwellian nature of the observed proton VDFs, numerical simulations will be needed to understand the dynamics of these distributions. Here, we used linear theory to estimate growth rates, dominant modes, and implied phase speeds. We acknowledge that a more sophisticated treatment could include additional nonlinear effects, such as finite-width trapping, as described in \cite{Karimabadi:1992} and applied to pick-up ions at Comet Halley in \cite{Karimabadi:1994}. However, such effects are beyond the scope of this paper and our observations report from a different heliospheric context.

These new PSP observations raise new inquiries on the contribution of these powerful ion beams to the solar-wind energy flux and the ongoing heating of the near-Sun solar wind. The source of the initial beam of field-aligned particles that eventually diffuses into the hammerhead region of velocity space also remains elusive. As PSP flies closer to the Sun, however, the ion VDFs will become more favorable to the SPAN-I field-of-view. More reliable measurements will lead to a better characterization of their complex shape as a function of plasma parameters of interest such as plasma $\beta$, general heliographic position, angular proximity to the heliospheric current sheet, and distance from magnetic topological changes. Thus, future PSP observations will likely provide answers to these important questions about the dynamics and thermodynamics of the inner heliosphere.

\acknowledgments

The SWEAP Investigation is supported by the PSP mission under NASA contract NNN06AA01C.
KGK is supported by NASA ECIP grant 80NSSC19K0912. D.V. is supported by STFC Ernest Rutherford Fellowship ST/P003826/1 and STFC Consolidated Grant ST/S000240/1. B.L.A. acknowledges NASA contract NNG10EK25C.

\bibliography{royalbib}



\end{document}